# Self-induced inverse spin Hall effect in permalloy at room temperature

Ayaka Tsukahara [#,1], Yuichiro Ando [#,1,2], Yuta Kitamura [#,1], Hiroyuki Emoto [1], Eiji Shikoh [1,$], Michael P. Delmo [1,&], Teruya Shinjo [1,2] and Masashi Shiraishi [1,2] *

1. Graduate School of Engineering Science, Osaka Univ., Toyonaka 560-8531, Japan.
2. Graduate School of Engineering, Department of Electronic Science and Engineering, Kyoto Univ., Kyoto 615-8530, Japan.

*corresponding author: Masashi Shiraishi (shiraishi@ee.es.osaka-u.ac.jp)
# These authors contributed equally to this study.
$ Present address: Department of Engineering, Osaka City Univ., Osaka 558-8585, Japan.
& Present address : Physics Department, De La Salle Univ., Manila 0922, Philippines.

**Inverse spin Hall effect (ISHE) allows the conversion of pure spin current into charge current in nonmagnetic materials (NM) due to spin-orbit interaction (SOI). In ferromagnetic materials (FM), SOI is known to contribute to anomalous Hall effect (AHE), anisotropic magnetoresistance (AMR), and other spin-dependent transport phenomena. However, SOI in FM has been ignored in ISHE studies in spintronic devices, and the possibility of "self-induced ISHE" in FM has never been explored until now. In this paper, we demonstrate the experimental verification of ISHE in FM. We found that the spin-pumping-induced spin current in permalloy (Py) film generates a transverse electromotive force (EMF) in the film itself, which results from the coupling of spin current and SOI in Py. The control experiments ruled out spin rectification effect and anomalous Nernst effect as the origin of the EMF.**

The interaction between spin and orbital angular momentum - the spin-orbit interaction (SOI) - generates anomalous Hall effect (AHE) [1,2], spin Hall effect (SHE) [3], inverse spin Hall effect (ISHE) [4] and other spin-related phenomena in condensed matter systems. Among these effects, the ISHE enables the conversion and detection of pure spin current - the flow of spin angular momentum without the flow of charge in a nonmagnetic material (NM) - into charge current. Strong SOI in NM allows the efficient conversion of pure spin current into charge current, which makes ISHE a valuable method for the characterization of spin injection and spin transport in various materials. SOI also exists in ferromagnetic materials (FM), which generates the AHE and anisotropic magnetoresistance (AMR). It is typical to use FM as a spin reservoir to generate spin current in NM. However, in the characterization of the spin-to-charge current conversion, only the ISHE in NM is considered and the SOI in FM is usually ignored. Nevertheless, we report in this paper that ISHE can be induced in FM, which results from the coupling of the spin current in FM and its own SOI.

The nickel (Ni)-iron (Fe) –$Ni_{80}Fe_{20}$- alloy, also known as permalloy (Py), is a widely used FM and spin reservoir for spin pumping. Spin pumping is a technique, which enables the magnetization dynamics in Py to be transferred, in the form of spin injection, in NM under a ferromagnetic resonance (FMR) condition [5-16]. The high permeability and weak magnetic anisotropy of Py, results in efficient spin injection and pure spin current generation in NM. Since, Py efficiently generates spin current in NM, it is highly possible that the spin current within the Py itself could effectively couple with its own SOI. It is known that Py exhibits AMR, which suggests that Py has a non-negligible SOI. Hence, in principle, a self-induced ISHE - the generation of ISHE in FM - from Py under FMR can manifest, although the experimental detection of this self-induced ISHE has never been reported, so far. Recently, Miao et al. reported the injection of pure spin current from yttrium-iron-garnet (YIG) in Py, resulting in ISHE in Py [17], where the flow of pure spin current in FM were also demonstrated. However, the pure spin current was generated by thermal gradient, which indicates that the observed effect was not the self-induced-ISHE of Py. In this paper, we demonstrate that the measured ISHE in Py in our experiments is the result of the self-generated pure spin current in Py that interacts with its own SOI, yielding an electromotive force (EMF). A number of control experiments ruled out other spurious effects that could generate EMF, such as the spin rectification effect (SRE) under irradiation of microwave [13, 18-20] and anomalous Nernst effect (ANE) [21-23].

Figure 1(a) shows the schematic structure of the devices. Three different types of substrates were used for device fabrication in this study; thermally oxidized $SiO_2$ (500 nm thick) / Si substrate, diamond substrate, and YIG (1 μm in thick) on Gadolinium-Gallium-Garnet (GGG) substrate. After surface cleaning of the substrates with acetone and 2-propanol, Py were deposited by electron beam evaporation at room temperature (RT). The dimension of the Py layer was 1.5 × 4.0 mm$^2$, and the thickness, $d$, of the Py layer was 10 nm. Two voltage leads separated by 2.5 mm gap for measuring electromotive force (EMF) were attached to the edge of the Py film with Ag paste. The sample was placed near the center of the $TE_{102}$ cavity, where the magnetic-field component of the microwave mode was the maximum and the electric-field component was minimum in the electron spin resonance (ESR) system (Bruker EMX10/12). Microwave mode with frequency, $f$, of 9.61±0.01 GHz, was applied to the samples. Static magnetic field, $H$, was applied to the Py layer at an angle, $θ_H$, as shown in Fig. 1(a). The direct current (DC) - EMF was measured by using a nanovoltmeter (KEITHLEY 2181A). For the AHE measurements, a Hall-bar-shaped Py layer ($d$ = 50 nm) with the dimension of 60 × 700 μm$^2$ was fabricated on thermally oxidized $SiO_2$/Si substrate by using electron beam lithography, lift-off process, and electron beam evaporation. A DC electric current ($I_x$ = 1 mA) and a perpendicular magnetic field, $H$, were applied in a Physical Properties Measurement System (Quantum Design) for detecting AMR. All measurements were performed at RT.

Figure 1(b) shows the ferromagnetic resonance (FMR) spectrum, $dI(H)/dH$ versus $H-H_{FMR}$, of the Py/$SiO_2$ sample, under excitation power, $P_{MW}$ = 200 mW, for $θ_H$ = 0°. FMR field, $H_{FMR}$, and FMR line width, $ΔW$, are estimated to be 115.6 mT and 7.8 mT. DC-EMF along y direction, $V_y$, as a function of $H-H_{FMR}$ is shown in Fig. 1(c). Interestingly, we measured a concomitant EMF signal (open circle in black) around $H_{FMR}$. Since, the device is composed only of Py layer and measuring electrodes, the induction of EMF signal is not trivial and unexpected. To be more quantitative, we apply a following fitting function to obtain the contribution of ISHE and AHE to the EMF in our device; [4]

$$V_y = V_{ISHE} \frac{\Gamma^2}{(H-H_{FMR})^2+\Gamma^2} + V_{AHE} \frac{-2\Gamma(H-H_{FMR})}{(H-H_{FMR})^2+\Gamma^2} \quad . \qquad … (1)$$

Here, $\Gamma$ denotes the damping constant. The first term, which has a symmetrical Lorentzian shape, corresponds to the contributions from the ISHE, whereas the second term, which has an asymmetrical shape, corresponds to that from the AHE (the detailed for confirming the ISHE is discussed later). We found that EMF is mainly induced by ISHE as indicated by the $V_{ISHE}$ fit in

Fig. 1(c) (the red solid line), where $V_{ISHE}$ = 19.7 μV, and the contribution of AHE to EMF is small, as indicated by $V_{AHE}$ fit (the blue dotted line). The ratio of $V_{ISHE}/V_{AHE}$ is estimated to be 7.1. We measured the magnitude of the $V_{ISHE}$ as a function of $\theta_H$, as shown in Fig. 1(d). The $\theta_H$ is varied from 0° to 360°. The EMF signal increases slightly as $\theta_H$ increases from 0° to 80°, but decreases sharply down to $V_{ISHE} \approx 0$ μV when $\theta_H$ approaches 90°. The polarity of $V_{ISHE}$ reverses to negative when $\theta_H$ exceeds 90°, and the increase of negative $V_{ISHE}$ proceeds further until $\theta_H$ = 100°, but then again decreases slightly when $\theta_H$ approaches 180°. This behavior is retraced back, like a mirror image, when $\theta_H$ is increased from 180° to 360°. This two-fold symmetry is roughly consistent with $\theta_H$ dependence of the ISHE-induced EMF in the previous spin pumping studies, $V_{ISHE} \sim J_s \times \sigma$, where $J_s$ and $\sigma$ are the spin current and spin angular momentum directions, respectively [4, 5-16]. The origin of slight discrepancy of peak and dip structures at $\theta_H$ =80°, 260°, and 100°, 280°, respectively, will be discussed later. The $V_{ISHE}$ as a function of $H-H_{FMR}$ for various $P_{MW}$ exhibits linear increase with the $P_{MW}$, (see solid circles in Fig. 1(e)), similar to those of other spin pumping devices [8-16]. The $\theta_H$ and $P_{MW}$ dependences of the measured $V_{ISHE}$ are consistent with ISHE-induced EMF appeared in spin pumping devices consisting of FM spin injector and NM spin channel. These results suggest that the ISHE can be induced in Py itself.

To understand how EMF could be induced by ISHE in Py, it is important to consider first how pure spin current is generated and flows in ferromagnetic layer. Since the polarity of the $V_{ISHE}$ is positive at $\theta_H$ = 0°, the spin current in the Py layer is expected to flow towards the Py/SiO$_2$ interface. At FMR condition, the magnetization precession is constantly maintained in the FM, and as a result, a part of the angular momentum of precessing local spin is transferred to conduction electrons, which polarizes its spin [6-8]. The spin polarization propagates as pure spin current into the NM if the NM is connected to the FM (the bilayer case). However, since Py is connected directly to SiO$_2$, it is unlikely that pure spin current would flow effectively in SiO$_2$ because it has very high electrical resistivity. Thus, if pure spin current can be generated, it should flow inside the Py layer and be converted to charge current by the ISHE because the Py is the only medium conductive. Since the spin relaxation in Py/SiO$_2$ interface is stronger than that in Py layer because of charged impurities on the SiO$_2$ (the spin scattering centers), it is possible that local spin damping at the interface induces a kind of spin density gradient in the Py that allows diffusive flow of pure spin current in the Py layer towards the interface. In fact, theory predicts that pure spin current can be generated in FM when non-uniform magnetization, i.e., spatial dependence of damping, exists [24]. We note that no spin current flows if the spin density in the Py layer is uniform. When spin scattering takes place at the $SiO_2$ by the charged impurities, resulting in the spin density gradient in the Py, the spin current flows into the Py layer towards the Py/$SiO_2$ interface. Here, the ISHE would be generated across the device, which possesses a symmetry represented by $V_{ISHE} \sim J_s \times \sigma$. This symmetry is corroborated by the polarity of the EMF and its reversals shown in Fig. 1(d).

If the above scenario is correct, a choice of substrates can verify this mechanism, because one can suppress the magnitude of the spin density gradient in Py layer by using a substrate with small spin damping. Thus, we selected first YIG (1 μm thick) on GGG as a substrate, because (1) YIG has an electrical resistivity comparable to $SiO_2$, (2) the magnetization damping of YIG is small (Gilbert damping constant [25], $\alpha \approx 6.7 \times 10^{-5}$), (3) $H_{FMR}$ of YIG is different from that of Py, and (4) the number of charged impurities on YIG is much less than that on $SiO_2$. It is noted that there was no spin angular momentum coupling between the YIG layer and the Py layer because surface treatments of the YIG before the Py deposition, which is necessary to realize effective coupling of spin angular momentum [26], was not carried out. In fact, no clear $V_{ISHE}$ was observed from the Py layer under FMR condition of the YIG layer. From above reasons, one can expect that the magnitude of the $V_{ISHE}$ of the Py layer on the YIG layer can be suppressed. The resulting FMR signal of the Py layer is shown in Fig. 2(a). The solid line shows the FMR signal for the Py/YIG sample, and the dotted line shows that for the Py/$SiO_2$ sample (for comparison). The Δ$W$ of the Py/YIG sample is estimated to be 3.4 mT, which is obviously smaller than that of the Py/$SiO_2$ sample. The smaller Δ$W$ indicates that spin dumping of the Py layer on the YIG layer is suppressed by changing the substrate from the $SiO_2$ to the YIG. Simultaneously, the $V_{ISHE}$ (the red line) for the Py/YIG sample is strongly suppressed as shown in the lower panel of Fig. 2(a). The magnitude of $V_{ISHE}$ for the Py/YIG sample is estimated to be 2.7 μV, which is one seventh of that observed in the Py/$SiO_2$ device (19.7 μV). We note that similar results were also observed in the Py/diamond (undoped) sample, as shown in Fig. 2(b). Undoped diamond is highly resistive, and has no magnetic damping and less charged impurities at the surface, similar to the case of YIG. The Δ$W$ ( 3.0 mT) and the $V_{ISHE}$ (-0.072 μV) of the Py on the diamond were successfully suppressed as expected. In addition, the reduction in $V_{ISHE}$ was observed by replacing Py with Cobalt (Co), whose $\theta_{SHE}$ and resistivity are expected to be small [27-29] (see Supplementary Materials (SM) A). These control experiments sufficiently support our claim that the observed $V_{ISHE}$ was attributed to the self-induced ISHE in the Py layer.

It is also clarified that the magnitude of the ISHE is changable by changing the spin dumping of adjacent materials or by the spin Hall angle of the FM layer.

There are several spurious effects, which can generate DC-EMF under the FMR condition, e.g., the ANE and the SRE due to the planar Hall effect (PHE). Here, we discuss the contribution of these effects on our experimental results. First, we focus on the ANE, which originates from microwave-induced temperature gradient in samples under the FMR condition. It is known that a microwave application to a sample induces thermal agitation and temperature increase. The temperature gradient, vertical to the film plane, generates charge current flow in the Py, yielding a lateral DC-EMF, perpendicular to the magnetization of the Py layer due to the ANE. In fact, $\theta_H$ dependence of the DC-EMF induced by the ANE exhibits the same behavior as that observed in this study. However, we note that the thermal conductivity, $\kappa$, for diamond, YIG and SiO$_2$ are 1,200, 6.0 and 1.2 Wm$^{-1}$K$^{-1}$, respectively, and that of air (adjacent of the surface of the Py layer) is 0.026 Wm$^{-1}$K$^{-1}$ [30-33]. Thus, if the ANE governs this effect, the DC-EMF from the Py/diamond sample should be the largest among the samples, because diamond behaves as an effective heat sink. However, the result is opposite, which strongly suggests that ANE can be ruled out. Furthermore, an additional control experiment was implemented as follows: we measured the $V_{ISHE}$ of the Py with various duration times of FMR. The duration time of the FMR condition in this study is defined as, $\Delta W$ [mT] / Sweeping rate[mT/s], where $\Delta W$ is the FMR line width, and the time was several seconds (the total measurement time was 2-20 min). We deduce that the sample cannot reach a steady state with thermal gradient in the sample. However, the magnitude of the EMF does not depend on the duration time, which also supports our claim (in more detail, see also SM B).

Hereafter, we discuss the contribution of SRE due to PHE, where the microwave-induced charge current generates DC-EMF.[13, 20]. First, we studied the SRE due to the AMR, to discuss the inductive current in the Py layer and measure the corresponding voltage from AMR, $V_{AMR}$, along the $x$ direction, as shown in Fig. 3(a).[13, 19] Typical $V_x$-$H$ curves of the Py/SiO$_2$ and the Py/YIG samples are shown in Fig. 3(b). Magnitude of the symmetrical EMF, which is $V_{AMR}$, as a function of $\theta_H$ for the Py/SiO$_2$ and the Py/YIG samples are displayed in Figs. 3(c) and 3(d), respectively. The almost similar behavior of $V_{AMR}$-$\theta_H$ curves, i.e., peaks and dips at $\theta_H$ = 80°, 260° and 100°, 280°, respectively, can be seen in Figs. 3(c) and (d). We theoretically calculated and estimated $V_{AMR}$, when $H$ is applied along the out-of-plane direction, and inductive current is generated along in-plane direction [13]. The inductive charge current density is expressed as, $J(t) = J_1 \cos(\omega t + \varphi)$, where $J_1$ is the magnitude of the inductive current and $\varphi$ is the phase angle between the rf magnetization and the rf current. The $V_{AMR}$ is given by $V_{AMR}(t) = w(\rho_0 + \rho_A \cos^2 \theta(t))(J(t))$. We assume that $\phi(t) = \phi(\theta_H)\cos(\omega t)$, where $\phi(t)$ is the angle between magnetization, $M$, and that of the equilibrium position, $M_0$, projected to the X-Z plane, $\phi(\theta_H)$ is the maximum angle of $\phi(t)$, i.e., the cone angle of the magnetization precession. Since $\theta(t) \cong \theta_M + \phi(t)$ and the magnetization precession angle is very small, $V_{AMR}(t)$ is given by

$$V_{AMR}(t) = w[\rho_0 + \rho_A \cos^2(\theta_M + \phi(t))](J_1 \cos(\omega t + \varphi))$$
$$\cong w[\rho_0 + \rho_A \cos^2 \theta_M - \rho_A \phi(\theta_H)\sin(2\theta_M)\cos(\omega t)](J_1 \cos(\omega t + \varphi)). \quad \ldots(2)$$

By taking the time average of the $V_{AMR}(t)$, we obtain the DC-EMF due to $V_{AMR}$, as follows:

$$V_{AMR} = -\frac{1}{2} w J_1 \rho_A \phi(\theta_H) \sin(2\theta_M) \cos\varphi. \quad \ldots(3)$$

Using $\theta_H$, the saturation magnetization, $M_s$, and $H$, the $\theta_M$ is expressed as, [34,35]

$$\frac{H}{4\pi M_s} \sin(\theta_H - \theta_M) = \sin\theta_M \cos\theta_M, \quad \ldots(4)$$

and the magnetization component along $X'$ direction, $m_{X'}$ is expressed as,[34, 35]

$$m_{X'}(\theta_H) = \frac{M_s h \gamma}{\alpha \sqrt{(4\pi M_s)^2 \gamma^2 \cos^4 \theta_M + 4\omega^2}}. \quad \ldots(5)$$

Here, $\gamma$ is the gyromagnetic ratio and $\omega = 2\pi f$, where $f$ is the microwave frequency. Since $\phi(\theta_H)$ is small, it is expressed as

$$\phi(\theta_H) \cong \tan\phi(\theta_H) = m_{X'}/M_s \quad \ldots(6)$$

Therefore, $V_{AMR}$ is expressed as,

$$V_{AMR} = -\frac{1}{2} w J_1 \rho_A \sin(2\theta_M) \cos\varphi \frac{h\gamma}{\alpha \sqrt{(4\pi M_s)^2 \gamma^2 \cos^4 \theta_M + 4\omega^2}}. \quad \ldots(7)$$

The result of the theoretically calculated $V_{AMR}$ as a function of $\theta_H$ is shown in Fig. 3(e). As can be seen, the theory nicely reproduces the experimental results, indicating that inductive current is actually generated.

Since, we clarified that non-negligible inductive current exists in the Py layer under FMR, the contribution of the SRE due to the PHE can be considered. In the same inductive current and the magnetic field configurations as those in the $V_{AMR}$ calculation, a DC component of the SRE due to the PHE, $V_{PHE}$, is expressed as:[34, 35]

$$V_{PHE} = -\frac{1}{2}wJ_1\rho_A cos\theta_M \frac{h\gamma\{2\alpha\omega cos\varphi-[4\pi M_s\gamma cos^2\theta_M+\sqrt{(4\pi M_s)^2\gamma^2 cos^4\theta_M+4\omega^2}]sin\varphi\}}{2\alpha\omega\sqrt{(4\pi M_s)^2\gamma^2 cos^4\theta_M+4\omega^2}}. \quad \ldots (8)$$

The experimentally observed $V_{ISHE}$ as a function of $\theta_H$ for the Py/YIG sample is shown in Fig. 4(a). For comparison, that of the Py/SiO$_2$ sample is once again displayed in Fig. 4(b). Since the PHE in the Py layer is not negligible as clarified in the above discussion, $\theta_H$ dependence of the $V_{PHE}$ is expected to be almost the same and irrelevant to a substrate type. However, the dependence shown in Figs. 4(a) and (b) is rather different. A result of theoretical calculation using eq. (8) is displayed in Fig. 4(c), where we set $\varphi$ =0.175° and $J_1$=3.0×10$^{10}$ A/m$^2$. While the theoretical result is quite similar with the result of the $V_{ISHE}$ in the Py/YIG sample, it is obviously different from the result in the Py/SiO$_2$ sample. Therefore, the non-negligible contribution of self-induced ISHE in Py should be considered in addition to the SRE. The $\theta_H$ dependence of the $V_{ISHE}$, due to the ISHE, is described by using the following equation, [34, 35]

$$V_{ISHE}(\theta_H) = |V_{ISHE}(\theta_H = 0^o)|J_s^{Nor}$$

$$= |V_{ISHE}(\theta_H = 0^o)|\frac{2\omega[4\pi M_s\gamma cos^2\theta_M+\sqrt{(4\pi M_s)^2\gamma^2 cos^4\theta_M+4\omega^2}]}{(4\pi M_s)^2\gamma^2 cos^4\theta_M+4\omega^2}. \quad \ldots(9)$$

Here, $J_s^{Nor} = J_s(\theta_M)/J_s(0^o)$ is the normalized spin current density, and the result is shown in Fig. 4(d). The theoretical results of the summation of the $V_{ISHE}$ and the $V_{PHE}$ as a function of $\theta_H$ for the Py/YIG and the Py/SiO$_2$ samples are shown in Figs. 4(e) and (f), respectively. We set $V_{ISHE}$= 6 and 19 μV, respectively. As can be seen, the theoretical result reproduces the experimental ones in great detail, strongly indicating that (1) there are two contributions to the EMF, and (2) the contribution of the ISHE is enhanced in the Py/SiO$_2$ sample comparing with that in the Py/YIG sample.

Finally, we estimated $\theta_{SHE}$ of Py by measuring the AHE of Py by using simple Hall measurement scheme (Fig. 5(a)). For the AHE measurements, a Hall-bar-shaped Py layer ($d$ = 50 nm) with the dimension of 60 × 700 μm$^2$ was fabricated on thermally oxidized Si substrate (SiO$_2$/Si) by using electron beam lithography, lift-off process, and electron beam evaporation. A DC electric current ($I_x$ = 1 mA) and a perpendicular magnetic field, $H$, were applied. Since the FMR measurements revealed that the magnetization of Py layer is aligned along perpendicular to the film plane at 1,000 mT, the linear relationship between the Hall voltage and $H$ above 1,000 mT can be ascribed to the normal Hall effect. Figure 5(b) shows the Hall voltage as a function of $H$ from -2,000 mT to 2,000 mT. The experimentally obtained Hall voltage is the open circles in black. The red line is the linear fit employed above ±1,000 mT, but extended down to 0 mT to obtain (via ordinate intercept) the Hall voltage, $V_y$ due to AHE in Py, which is found to be 19.2 μV. The anomalous Hall resistivity [36, 37] is $\rho_{AHE} = \frac{V_y d}{I_x} = \sigma_{AHE}\rho^2$, where $\rho$ is the resistivity under the zero magnetic field, $\sigma_{AHE}$ is the anomalous Hall conductivity, and $I_x$ is the applied dc current. The relationship between the conductivities of the spin Hall effect and the AHE is described as $\sigma_{SHE} = (\frac{1}{P})\sigma_{AHE}$, where $P$ is the spin polarization of Py. The $P$ of Py is typically 0.2 - 0.5 [38, 39], and the $\rho$ and the $\rho_{AHE}$ were experimentally estimated to be $3.62 \times 10^{-7}$ Ω m and $9.6 \times 10^{-10}$ Ω m, respectively. Thus, the $\theta_{SHE}(= \frac{\sigma_{SHE}}{\sigma_F})$ was calculated to be 0.005 - 0.013, which is almost the same magnitude as that of Pd, the metal often used for ISHE studies [35].

In summary, we experimentally demonstrated the manifestation of the self-induced ISHE in Py. The conversion efficiency in FM is comparable to that of known noble metals, which can be used in various applications.

**Acknowledgements**

This research was partly supported by a Grant-in-Aid for Scientific Research from the MEXT, Japan and by the Global COE program of "Core Research and Engineering of Advanced Materials Interdisciplinary Education Center for Materials Research".

**Figure captions**

Figure 1

(a) A schematic illustration of the devices. The dimension of the $Ni_{80}Fe_{20}$ (Py) layer is 1.5 mm (*l*) × 4.0 mm and the thickness, *d*, is 10 nm. Two electrodes are attached on the Py layer by using Ag paste. Electrodes separation width, *w*, is 2.5 mm. The static external magnetic field, *H*, is applied at an angle of $\theta_H$ to the Py film plane. (b) FMR spectra, d*I*(*H*)/d*H*, of the Py/SiO$_2$ device for $\theta_H$ = 0°, as a function of *H*-*H*$_{FMR}$, where *I* is microwave absorption intensity in arbitrary unit. The microwave power is 200 mW, and the ferromagnetic resonance field, $H_{FMR}$ and peak-to-peak width, $\Delta W$, of the FMR signal is estimated to be 115.6 mT and 7.8 mT, respectively. (c) The *H* dependences the electromotive force (EMF), $V_y$, for $\theta_H$=0°. (d) $\theta_H$ dependence of the $V_{ISHE}$. The microwave power is 200 mW. Background signals are subtracted from the raw data by taking an average of the signals at $\theta_H$=0° and 180°, and so on. (e) Microwave power, $P_{MW}$, dependence of the $V_{ISHE}$ and the $V_{AHE}$, where the $V_{ISHE}$ and $V_{AHE}$ are the amplitudes of the electromotive forces due to the ISHE and the AHE, respectively. The red and blue solid lines are results of linear fitting of the experimental data. The inset shows *H* dependence of the $V_y$ under different microwave excitation powers when $\theta_H$ = 0°.

Figure 2

(a) The *H* dependence of the d*I*(*H*)/d*H*, when $\theta_H$=0° for Py/YIG sample (solid line). The FMR signal for Py/SiO$_2$ sample is also displayed as black broken line. The procedures of the measurements of EMFs and the geometry of the sample are the same as those of the Py/SiO$_2$ sample. $H_{FMR}$ and $\Delta W$ are estimated to be 107.3 mT and 3.4 mT, respectively. The *H* dependence of the electromotive force, $V_y$, measured for the Py/YIG, when $\theta_H$ = 0°. The open circles are the experimental data and the red and blue solid lines are the contribution from the ISHE and the AHE, respectively. The notable point is that the EMF by the ISHE from this sample is quite small and the $V_{ISHE}$ was estimated to be 2.7 μV.

Figure 3

(a) A schematic illustration of the samples for measurement of the SRE due to the AMR. (b) Typical $V_x$-*H* curves for the Py/SiO$_2$ and Py/YIG samples. The microwave power is 200 mW. $\theta_H$ dependence of the $V_{AMR}$ for the (c) Py/SiO$_2$ and (d) Py/YIG samples. (e)

Theoretically calculated $V_{AMR}$ as a function of $\theta_H$. The fitting parameters are $\alpha$ = 0.0096, $M_s$ = 71.9 mT, $\varphi$ = 0.175, $h_{rf}$ = 0.06 mT, $f$ = 9.622 GHz, $\gamma$ = 1.86×10$^{11}$ T$^{-1}$s$^{-1}$, $\rho_A$ = 7.24×10$^{-9}$ Ωm, and the current density for $V_{AMR}$ was set to be 1.1×10$^9$ A/m$^2$.

Figure 4

$\theta_H$ dependence of the $V_{ISHE}$ for (a) Py/SiO$_2$ and (b) Py/YIG samples. The microwave power is 200 mW. Theoretically calculated (c) $V_{PHE}$ and (d) $V_{ISHE}$ as a function of $\theta_H$. $V_{PHE}+V_{ISHE}$ as a function of $\theta_H$, where the magnitude of $V_{ISHE}$ are (e) 6 and (f) 19 μV. The fitting parameters are $\alpha$ = 0.0096, $M_s$ = 71.9 mT, $\varphi$ = 0.175, $h_{rf}$ = 0.06 mT, $f$ = 9.622 GHz, $\gamma$ = 1.86×10$^{11}$ T$^{-1}$s$^{-1}$, $\rho_A$ = 7.24×10$^{-9}$ Ωm, and the current density for $V_{PHE}$ was set to be 2.2×10$^{10}$ A/m$^2$.

Figure 5

(a) A schematic illustration of the Py/SiO$_2$ sample for Hall effect measurement. The dimension of the Py layer was 60 × 700 μm$^2$ and the thickness, $d$, was set to be 50 nm. A dc current, $I_x$, of 1 mA and a perpendicular magnetic field, $H$, are applied. (b) The $H$ dependence of the Hall voltage of the Py/SiO$_2$ sample. The linear relationship between the Hall voltages and the $|H|$ above 1 T is reasonably explained by the normal Hall effect and the red line is a linear fit of the data at $|H|$ > 1 T. The anomalous Hall voltage of the Py layer is obtained from the ordinate intercept of the fitting line.

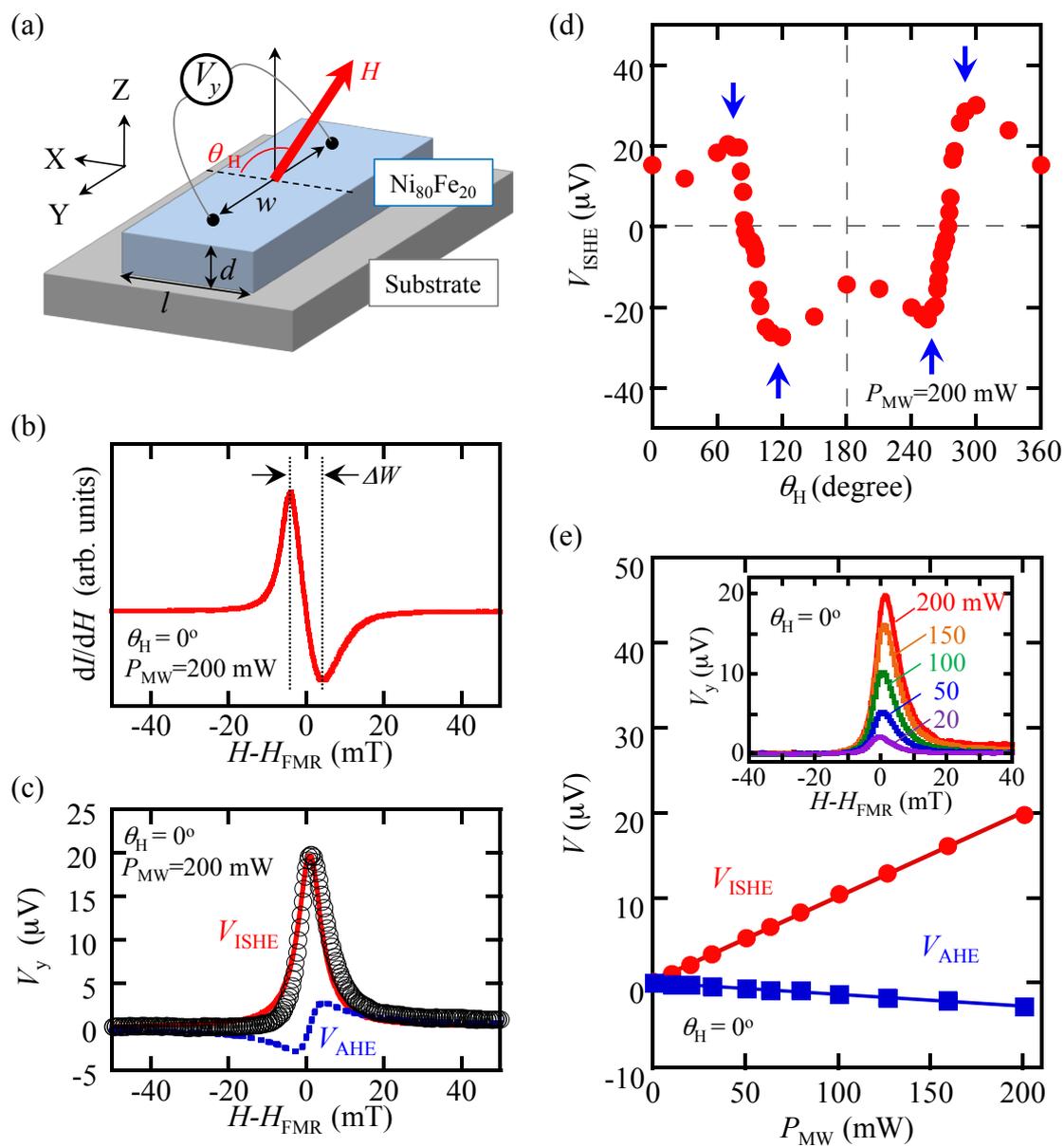

Fig. 1 A. Tsukahara et al.

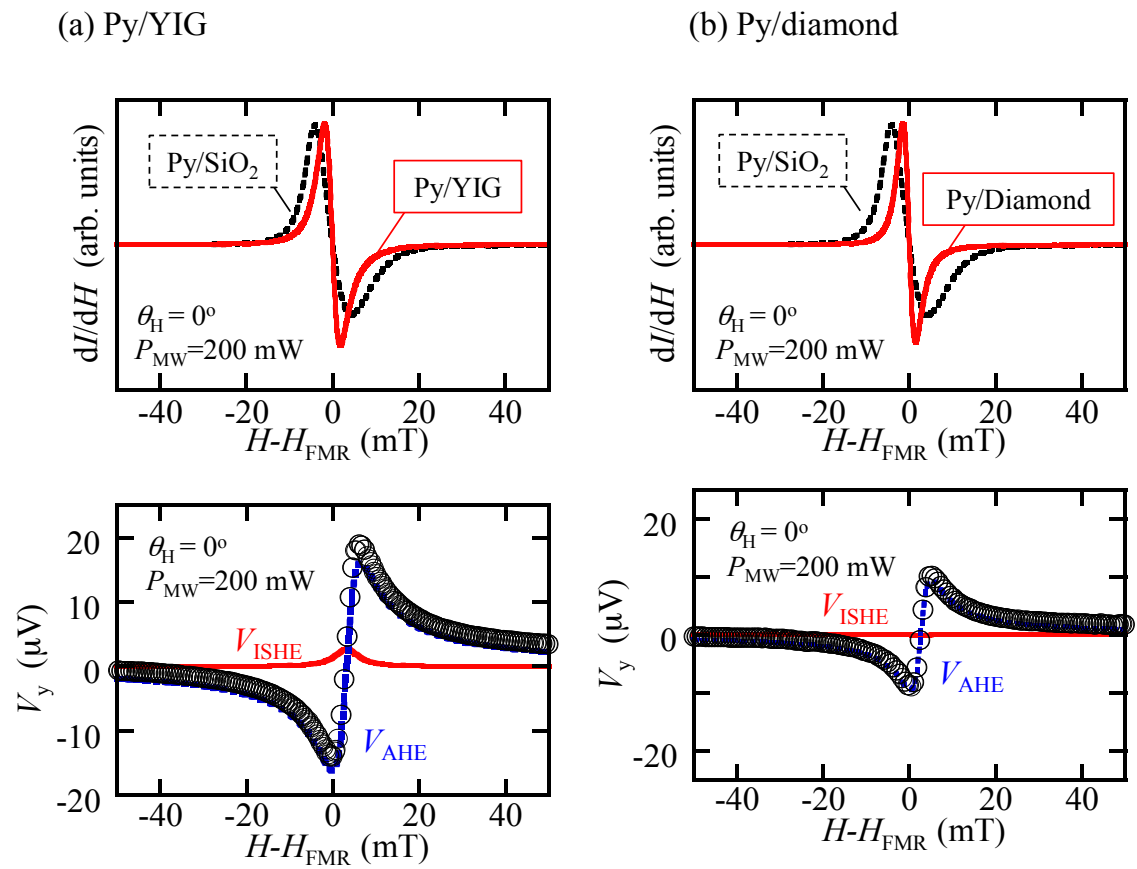

Fig.2 A. Tsukahara et al.

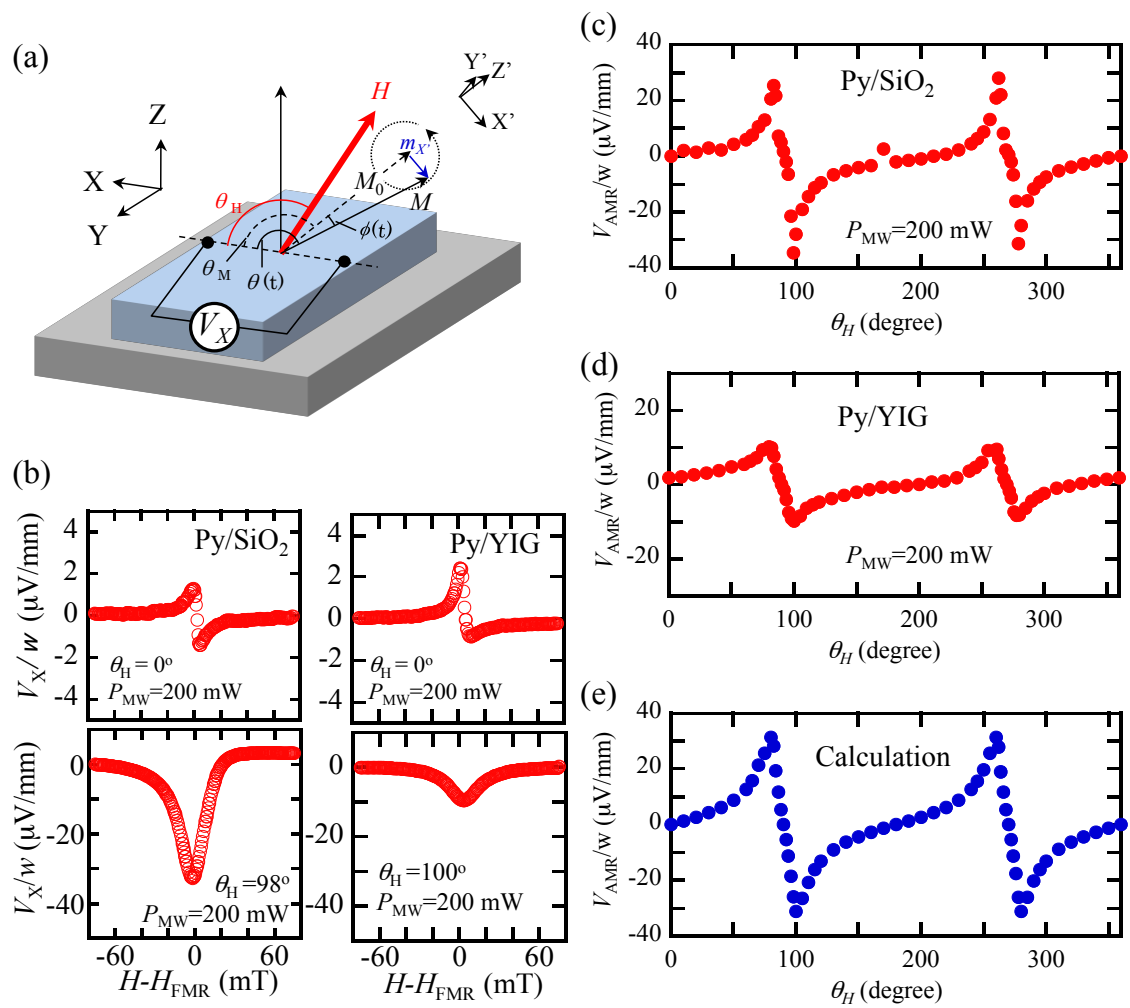

Fig. 3 A. Tsukahara et al.,

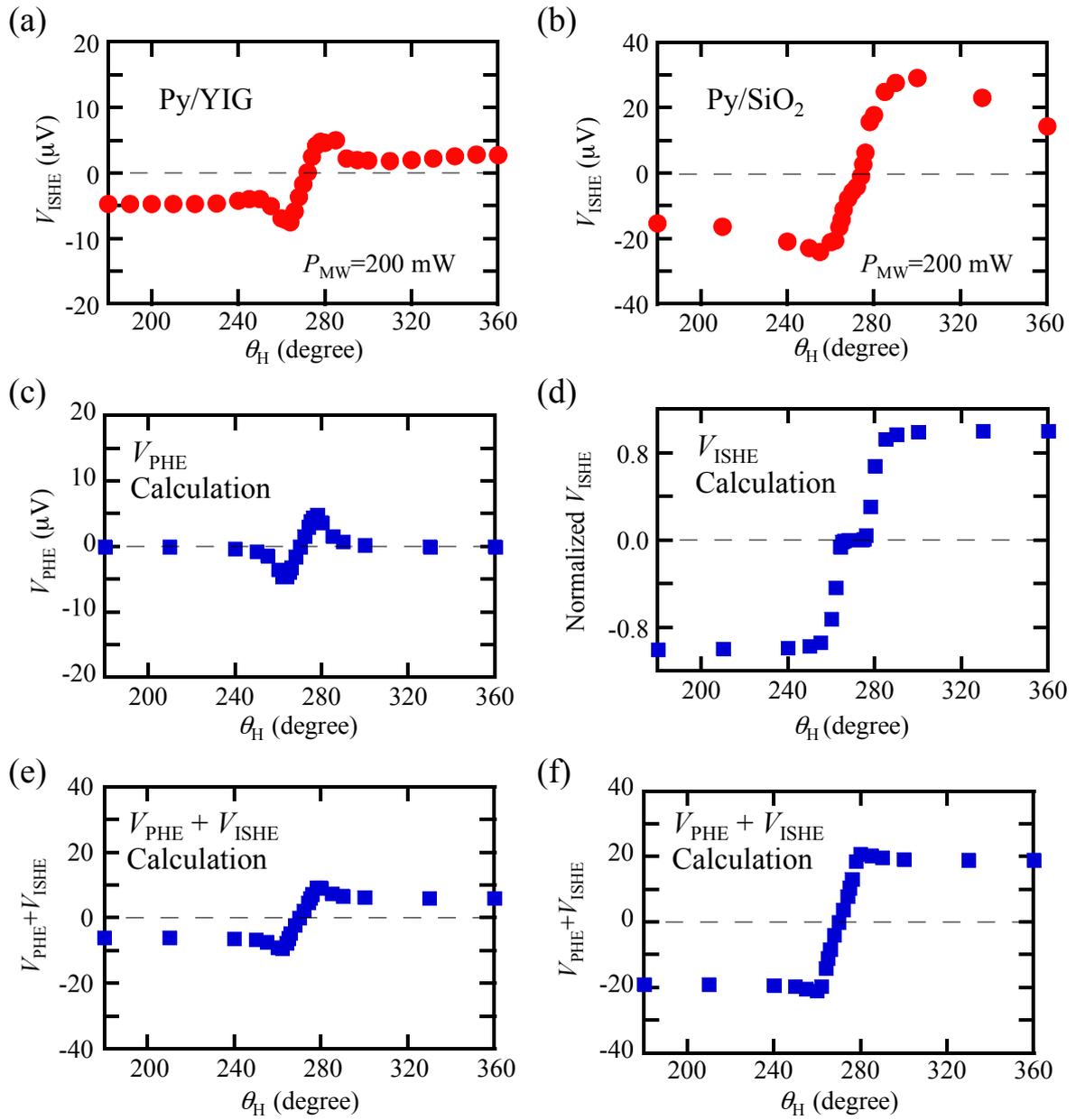

Fig.4 A. Tsukahara et al.

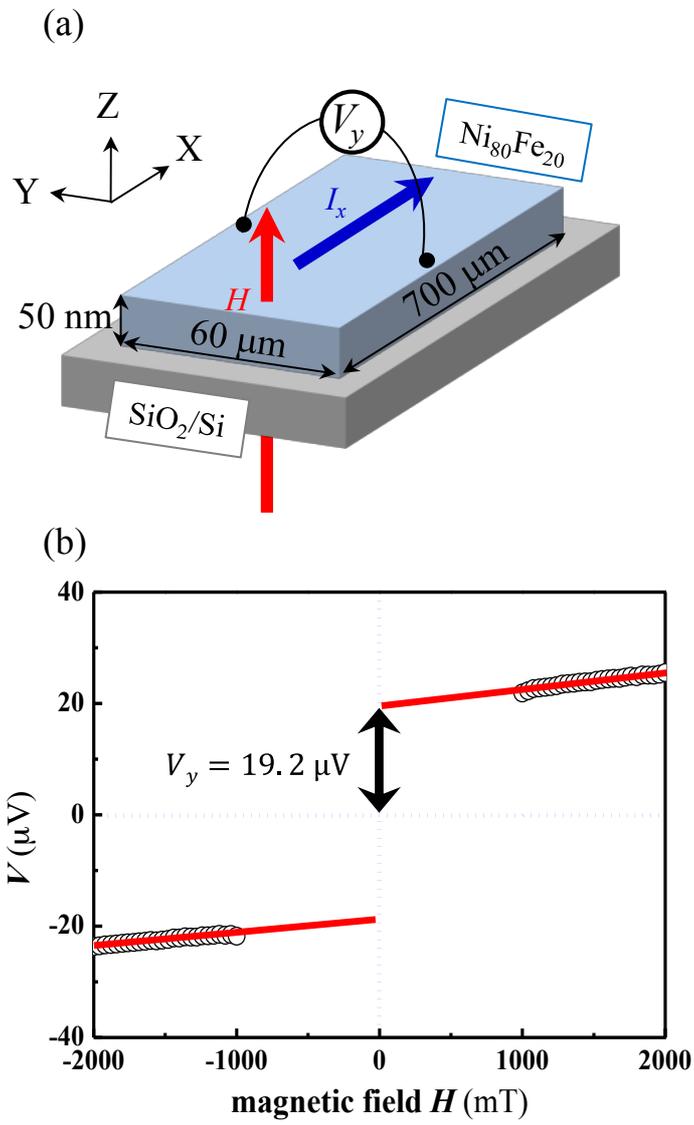

Fig.5 A. Tsukahara et al.